# Lossy Data Compression Using Logarithm

**Vivek Kumar, Srijita Barthwal, Rishabh Kishore, Ruchika Saklani, Anuj Sharma, Sandeep Sharma**
Department of Electronics and Communication Engineering
Dehradun Institute of Technology, Dehradun, Uttarakhand

**Abstract:** Lossy compression algorithms take advantage of the inherent limitations of the human eye and discard information that cannot be seen. [1] In the present paper a technique termed as Lossy Data Compression using Logarithm (LDCL) is proposed to compress incoming binary data in the form of a resultant matrix containing the logarithmic values of different chosen numeric sets. The proposed method is able to achieve compression ratio up to 60 in many major cases.

## I. Introduction

When we speak of a compression algorithm we are actually referring to two algorithms. There is the compression algorithm that takes an input X and generates a representation $X_c$ that requires fewer bits, and there is a reconstruction algorithm that operates on the compressed representation $X_c$ to generate the reconstruction Y. These operations are shown schematically in Figure 1. Based on the requirements of reconstruction, data compression schemes can be divided into two broad categories lossless compression schemes, in which Y is identical to X, and lossy compression schemes, which generally provide much higher compression than lossless compression but allow Y to be different from X. [2]

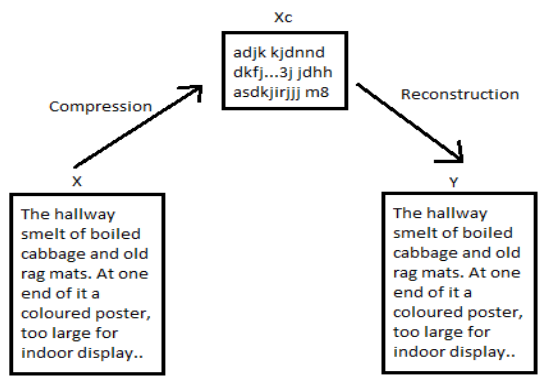

**Figure 1:** Compression and Reconstruction

Straightforward approaches for scientific data compression exist in lossless techniques designed specifically for floating-point data. However, due to the high variability of the representation of floating-point numbers at the hardware level, the compression factors realized by these schemes are often very modest [5,8]. Since most post-run analysis is robust in the presence of some degree of error, it is possible to employ lossy compression techniques rather than lossless, which are capable of achieving much higher compression rates at the cost of a small amount of re-construction error. As a result, a number of approaches have been investigated for lossy compression of scientific simulation datasets including classical [7] and diffusion wavelets [4], spectral methods [6], and methods based on the techniques used for transmission of HDTV signals [3]. However, these approaches are either applicable only to simulations performed on structured grids or have high computational requirements for in situ data compression applications. [9]

Lossy compression techniques involve some loss of information, and data that have been compressed using such techniques generally cannot be recovered or reconstructed exactly. In return for accepting this distortion in the reconstruction, we can generally obtain much higher compression ratios than is possible with lossless compression. [2]

The current paper is based on lossy data compression technique providing a much higher compression ratio and suitable for compressing multimedia and other related extensions. The paper is further divided into five categories. We premise our approach in section II. Section III sparks light upon the experimental design and results. Furthermore, section IV & V derives the conclusion and future scope of the LDCL technique respectively. The paper ends in section VI highlighting the reference of the work.

## II. Principle Operation

In the present technique, Lossy Data Compression using Logarithm (LDCL), the compression and reconstruction of the data is achieved in a hierarchy of following steps defined for each compression and reconstruction respectively.

### II.I Compression:

**Mapped Sequence** Binary sequence of the stored data is taken as an input and four known pairs (00, 01, 10, and 11) of 0 and 1 are mapped to their pre-defined respective numbers (2, 3, 4, and 5). If the sequence came out to be odd then "1" is added in front of the sequence and a flag is pointed to active high along with the resultant matrix of log values. Given below is a binary sequence and corresponding mapped sequence.

11010010111111111100…………. .. **(i)**
| | | | | | | | | |
5  3 2 4  5 5 5 5 5  2 ………….. .. **(ii)**

**Redundant Mapped Sequence** Once a mapped sequence is obtained; now check for the consecutive repetition of numbers. If no number repeats itself for more than four times at an instant then proceed ahead, if in case, any number repeats more than four times consecutively then arrange those numbers in a format specified as: "**1xn1**", where '**x**' represents the number itself and "**n**" represents the number of iterations involved.

Taking sequence of equation (ii):
Mapped Sequence:
    5324555552…………..
Redundant Mapped Sequence:
    5324**1551**2………….. ..**(iii)**

**Set Decomposition** Since, the redundant mapped sequence is holding known number of integers into it. So, certain numbers of sets with fixed number of integers are formed. The threshold to the number of integers in a set can be defined as per the hardware and software system capabilities. We take threshold to be as 300 integers per set. However, less the threshold lesser the error we obtain.

**Subtraction** Once all the sets are framed, choose a *default number* having all integers being "9" and number of digits being equal to the number of digits minus one chosen for each sets. For instance, we take 300 digits for each set, so, our *default number* should be containing 299 digits.

Now, start subtracting each set with the default number and finally a matrix of order 2*n is obtained where "n" represents total number of sets made. The row one of the matrix consists of the residue numbers left after subtraction and row two of the matrix consists of the number of times the subtraction takes place for each "n" sets.

$$U = \begin{matrix} s1 & s2 & \ldots\ldots\ldots\ldots & sn \\ m1 & m2 & \ldots\ldots\ldots\ldots & mn \end{matrix} \quad ..\textbf{(iv)}$$

**Logarithm of the matrix** Proceeding with the logarithm, the mandatory step to be acknowledged is what should be the "base" of the logarithm? Here, the base must be equal to the number chosen for the subtraction i.e. the default number. Now, take the logarithm of each element of the matrix given in equation (iv). Finally, a matrix of order 2*n with the logarithm of equation (iv) is obtained which depicts the desired compressed data.

### II.II Reconstruction

The reconstruction of the lossy compressed data is achieved by progressing with the anti-steps of the compressed steps.

**Antilog of the matrix** The very first step of the reconstruction includes taking the antilog of the given matrix with base familiar to the reconstruction end beforehand to be as chosen default number.

After taking the antilog, the resultant matrix would be having numbers which are approximately equivalent to those as achieved in the matrix of equation (iv). This is the very step which is solely responsible for the loss of data due to the inexactness of the returned values.

**Addition** Once, the matrix is obtained, start adding the default number to the numbers of row one with the number of times the number allotted to each elements of same column. This will produce the "n" number of sets.

**Forming the titan number** Collect all sets together in a sequential manner to form the final giant number.

**Non-Redundant Mapped Sequence** From the achieved giant number seek for the format "**1xn1**" and replace it with the number "x" with "n" number of times.

**Non-Mapped Sequence** Follow the reverse mapping sequence, now by mapping the numbers 2, 3, 4 and 5 with their corresponding binary pairs 00, 01, 10 and 11 respectively. Check for the active high flag, if detected an active high flag then remove "1" from the start of the sequence. This will provide us the original sequence before compression took place.

## III. Experimentation Design and Results

**Evaluation Methodology** To analyze the performance measure of the proposed technique, the compression effectiveness was measured by compression ratio (CR) which is defined as

$$CR = \frac{\text{Compressed size}}{\text{Decompressed size}} \quad ..(v)$$

Furthermore, to calculate the error introduced by the lossy compression we took root mean squared error (RMSE) which is defined as:

$$RMSE = \sqrt{1/|V|} \sum_{i=1}^{|V|} |l_j - t_j|^2 \quad ..(vi)$$

where $l_j$ is the original value of the set before taking logarithm and $t_j$ is their resultant reconstructed sets after taking respective antilog of each sets.

The experiments were performed in Matlab®.

**Results** Table 1 illustrates the performance measure of the technique based on the compression ratio for 5 different conducted experiments. In each experiment different numbers of sets are achieved after subtraction with the default number (A number 299 digits long of which all integers are "9".) and later performing logarithmic operation.

**Table 1:** Compression Ratio results.

| Exp. | Original Size (in bytes) | Compressed Size (in bytes) | CR |
|---|---|---|---|
| 1. | 1,073,741,824 | 57,266,240 | 18.74 |
| 2. | 3,221,225,472 | 53,687,091 | 60.00 |
| 3. | 5,368,709,120 | 143,165,576 | 37.50 |
| 4. | 8,589,934,592 | 257,698,037 | 33.33 |
| 5. | 10,073,741,824 | 178,956,970 | 56.29 |

The RMSE error corresponding to each experiment is shown in table 2.

**Table 2:** RMSE Results

| Experiment | RMSE |
|---|---|
| 1. | 0.20E+190 |
| 2. | 1.80E+296 |
| 3. | 0.94E+204 |
| 4. | 0.56E+221 |
| 5. | 1.56E+257 |

Hence, analyses of table 1 & 2 prove that larger the threshold default number larger will be the RMSE. Therefore, a threshold should be set according to the type of the compression demanded. If less size is demanded with the compromise of loss of some information then higher default number is a must choice however, if information must be retained to an extent then small default number is to be taken.

## IV. Conclusion

In this paper, we introduced a paradigm of lossy compression of data by taking the binary sequence and decomposing it into sets after mapping and redundant mapping to produce a matrix of logarithm values which is finally stored as compressed data. Our comprehensive set of experiments showed that this algorithm achieve compression which results in storage requirements that on average, are able to achieve compression ratio up to 60. The LDCL is quite useful to compress multimedia files into vary less size if taken large default number and simultaneously it is useful to compress text files with small default number.

## V. Future Scope

The next stage of the LDCL demands the reduction in the value of RMSE with large default number. Therefore, to make the technique more efficient an error approximation method will be introduced that will approximate the resultant values more closely to the original ones.